\documentclass[a4paper,11pt]{article}
\title{Markov property and  
strong additivity of von Neumann entropy 
 for graded quantum systems}
\date{}
\author{Hajime Moriya }
\usepackage{latexsym}%
\usepackage{amssymb} %
\usepackage{amsmath} %
\usepackage{amsthm} %
\usepackage{amscd} %
\newcommand{\R} {{\mathbb{R}}}%
\newcommand{\Com} {{\mathbb{C}}}%
\newcommand{\NN}{{\mathbb{N}}}%
\newcommand{\vp}{\varphi}
\newcommand{\lam}{\lambda}%
\newcommand{\ome}{\omega}
\newcommand{\Fin}{{\mathfrak F}}%
\newcommand{\I}{{\mathrm{I}}}%
\newcommand{\J}{{\mathrm{J}}}%
\newcommand{\IuJ}{\I \cup \J}%
\newcommand{\Tr}{\mathbf{Tr}}%
\newcommand{\tr}{\mathbf{tr}}%
\newcommand{\Al}{{\cal{A}}}%
\newcommand{\Bl}{{\cal{B}}}%
\newcommand{\Ale}{\Al^{e}}%
\newcommand{\Alo}{\Al^{o}}%
\newcommand{\AlI}{\Al_{\I}}%
\newcommand{\AlJ}{\Al_{\J}}%
\newcommand{\AlIuJ}{\Al_{\IuJ}}%
\newcommand{\AlIe}{\AlI^{e}}%
\newcommand{\AlIo}{\AlI^{o}}%
\newcommand{\AlJe}{\AlJ^{e}}%
\newcommand{\AlJo}{\AlJ^{o}}%
\newcommand{\Ala}{\Al_{A}}%
\newcommand{\Alb}{\Al_{B}}%
\newcommand{\Alc}{\Al_{C}}%
\newcommand{\Alab}{\Al_{AB}}%
\newcommand{\Albc}{\Al_{BC}}%
\newcommand{\Alac}{\Al_{AC}}%
\newcommand{\Alabc}{\Al_{ABC}}%
\newcommand{\spa}{\bar{A}}%
\newcommand{\spb}{\bar{B}}%
\newcommand{\spc}{\bar{C}}%
\newcommand{\Alsa}{\Al^{s}_{A}}%
\newcommand{\Alsb}{\Al^{s}_{B}}%
\newcommand{\Alsc}{\Al^{s}_{C}}%
\newcommand{\Alscac}{\Al^{\tilde{s}}_{C}}%
\newcommand{\Alsab}{\Al^{s}_{AB}}%
\newcommand{\Alsbc}{\Al^{s}_{BC}}%
\newcommand{\Alsz}{\Al^{s}_{ABC}}%
\newcommand{\Alz}{\Alabc}%
\newcommand{\Alae}{\Ala^{e}}%
\newcommand{\Albe}{\Alb^{e}}%
\newcommand{\Alce}{\Alc^{e}}%
\newcommand{\Albce}{\Albc^{e}}%
\newcommand{\Alao}{\Ala^{o}}%
\newcommand{\Albo}{\Alb^{o}}%
\newcommand{\Alco}{\Alc^{o}}%
\newcommand{\Albco}{\Albc^{o}}%
\newcommand{\id}{{\mathbf{1} } }
\newcommand{\qedb}{\hbox{\rule[-2pt]{3pt}{6pt}}}%
\newcommand{\proofend}{{\hfill \qedb}}
\newcommand{\nonum}{\nonumber}%
\newcommand{\Ezab}{E^{A,B,C}_{A,B} }%
\newcommand{\Eab}{E_{A,B} }%
\newcommand{\Ebc}{E_{B,C} }%
\newcommand{\Ea}{E_{A} }%
\newcommand{\Eb}{E_{B} }%
\newcommand{\Eaba}{E^{A,B}_{A} }%
\newcommand{\Eabb}{E^{A,B}_{B} }%
\newcommand{\Ebcb}{E^{B,C}_{B} }%
\newcommand{\EBl}{E^{\Al}_{\Bl} }%
\newcommand{\Tsharp}{T^{\sharp} }%
\newcommand{\rhovp}{\rho_\vp}%
\newcommand{\rhoaclam}{\rho_{AC, \lambda} }%
\newcommand{\psiz}{\psi_{ABC }}%
\newcommand{\psiab}{\psi_{AB }}%
\newcommand{\psibc}{\psi_{BC }}%
\newcommand{\psic}{\psi_{C}}%
\newcommand{\psib}{\psi_{B}}%
\newcommand{\psia}{\psi_{A}}%
\newcommand{\psiac}{\psi_{AC }}%
\newcommand{\omeac}{\ome_{AC }}%
\newcommand{\rhopsibc}{\rho_{\psibc}}%
\newcommand{\rhopsib}{\rho_{\psib}}%
\newcommand{\rhopsia}{\rho_{\psia}}%
\newcommand{\rhopsic}{\rho_{\psic}}%
\newcommand{\rhopsiab}{\rho_{\psiab}}%
\newcommand{\rhopsiac}{\rho_{\psiac}}%
\newcommand{\rhopsiz}{\rho_{\psiz}}%
\newcommand{\rhopsiae}{{\rhopsia}_{+}}%
\newcommand{\rhopsiao}{{\rhopsia}_{-}}%
\newcommand{\rhopsice}{{\rhopsic}_{+}}%
\newcommand{\rhopsico}{{\rhopsic}_{-}}%
\newcommand{\sh}{\widehat{S}}%
\newcommand{\TrI}{\Tr_{\I}}%
\newcommand{\TrJ}{\Tr_{\J}}%
\newcommand{\TrIuJ}{\Tr_{\IuJ}}%
\newcommand{\Hil}{{\cal H}}%
\newcommand{\Hilspa}{\Hil_{\spa}}%
\newcommand{\Hilspb}{\Hil_{\spb}}%
\newcommand{\Hilspc}{\Hil_{\spc}}%
\newcommand{\aicr}{a_i^{\ast}}%
\newcommand{\ai}{a_i}%
\newcommand{\vi}{v_{i}}%
\newcommand{\vI}{v_{\I}}%
\newcommand{\vJ}{v_{\J}}%
\newcommand{\vIJ}{v_{\IuJ}}%
\newcommand{\vA}{v_{A}}%
\newcommand{\va}{v_{A}}%
\newcommand{\vb}{v_{B}}%
\newcommand{\vab}{v_{A, B}}%
\newcommand{\vpBl}{\vp_{\Bl}}%
\newcommand{\Bex}{\Al \cap {\Bl}^{\prime}}%
\newcommand{\gammat}{\gamma_{\theta}}
\newcommand{\lami}{\lambda_i}%
\newcommand{\omeaci}{\ome_{AC,i}}
\newcommand{\omebi}{\ome_{B,i}}
\newcommand{\omeabc}{\ome_{ABC}}
\newcommand{\cev}{c_{+}}
\newcommand{\cod}{c_{-}}
\newcommand{\aev}{a_{+}}
\newcommand{\aod}{a_{-}}
\newcommand{\rhoone}{\rho_{1}}
\newcommand{\rhotwo}{\rho_{2}}
\begin{document}
\maketitle
\theoremstyle{plain}%
\newtheorem{thm}{Theorem}%
\newtheorem{cor}[thm]{Corollary}%
\newtheorem{lem}[thm]{Lemma}%
\newtheorem{pro}[thm]{Proposition}%
\begin{abstract}
It is easily verified that the  quantum Markov property 
is  equivalent to 
 the  strong additivity of von Neumann entropy 
for graded quantum systems.   
However,  the structure of Markov states for 
  graded systems is different from that 
 for   tensor-product systems which have  trivial grading.
  For  three-composed graded systems 
we have   U(1)-gauge invariant Markov states 
whose restriction to   the  
 marginal pair of subsystems is  nonseparable.
\end{abstract}
\section{Introduction}
\label{sec:INTRO}
We are interested in characterization of state correlations
 for general  composite systems which do  not necessarily 
satisfy the local commutativity.  We  
  consider   specifically 
finite-dimensional  quantum  systems   
with  graded commutation relations.
For such nonindependent case  as  well, 
 the total system
can  be divided  into  subsystems  on 
 disjoint  subregions  and 
 the notion of state correlation   
 among them makes sense.

It has been noted  that 
 known criterions of separability  for tensor-product 
 systems   should be altered for   lattice fermion systems \cite{SEP}
 when   fermion hopping terms   are  present.
(Note that   purely fermionic correlation  due to 
 hopping terms cannot be  distilled.)  

We are going to discuss
quantum Markov  property  \cite{ORI},    
 a quantum version of Markovness invented  
by Accardi. This is given by means of    quasi-conditional expectations 
and has  played    various  roles, see, e.g., \cite{TOP}.
We can view the  Markov property  
as a   kind of characterization   of  
state correlation for composite systems.
A  pivotal  example of  quantum composite   systems 
 is   tensor product of 
  Hilbert spaces for which lots of works,  
prominently those on    
 Markov   chains for 
one-dimensional quantum spin lattice systems  have been done.  
We  should   note  that
  the  definition
 of  Markov property   
has been  given under a very general setting
that  is  not limited  to 
the  tensor-product case.
Namely, it does not require  in principle  any specific
 {\it{algebraic 
 location}}  among subsystems imbedded in the total
system \cite{CEC}. 
Its   detailed  analysis   
for  nonindependent  systems, however,    has 
 started 
only recently.
\cite{CARacc} has investigated  Markov chains 
for  one-dimensional (spinless) fermion 
lattice systems and    clarified   that 
 the  notion of  Markov property  
and  of the Markov chain
is  well applicable to  such  fermionic case.
(More precisely, the above Markov chain refers to 
(d)-Markovian chain  \cite{ORI}, 
see  also \cite{OHNO} on the generalized Markov chain.)
 Furthermore,  a   class of  U(1)-gauge invariant Markov chains  
 with   fermionic hopping correlations
has been  given. 

It has been shown   that the Markov property 
 is tightly related to the 
sufficiency of conditional expectations 
 through the strong subadditivity of von Neumann
 entropy: A state of a  three-composed tensor-product  system
is  markovian if and only if 
it takes the equality 
for  the  strong subadditivity  inequality  of  entropy,   
which will be referred to as 
`the strong additivity of entropy'  \cite{PETZMONO} \cite{HAYDEN} 
\cite{MOSO} \cite{JEN}. 

We   show that   a  
similar equivalence relation of  the Markov property 
 and the strong additivity of entropy
is valid  for  graded quantum systems.
Its    proof proceeds in much the same way 
  as that  for the  tensor-product case  
following   \cite{PETZQUART}
  (whose   methods and results  can be used  in the present 
nonindependent situation)
  with   some  simple modifications  due to the   grading.

We  now introduce  the graded systems under our consideration.
Let $\Fin$ be a lattice and
  $\{\AlI; \  \I \in \Fin\}$ be a family of $\ast$-algebras
that  have a common unital element    denoted $\id$.
 If $\I\subset \J$, then $\AlI\subset \AlJ$,
and if $\I\cap\J=\emptyset$, then 
 $\AlI\cap \AlJ=\Com \id$. 
Let
 $\Theta$ be an   involutive $\ast$-automorphism of $\Al$
that determines  the  grading  as  
\begin{eqnarray}
\label{eq:EO}
 \Ale := \{A \in \Al \; \bigl| \;   \Theta(A)=A  \},\quad  
 \Alo := \{A \in \Al  \; \bigl| \;  \Theta(A)=-A  \}. 
 \end{eqnarray}
We assume that 
 our  grading transformation $\Theta$ is nontrivial.
The above   $\Ale$ and $\Alo$ (which is not empty) are 
called the even and  odd parts  of $\Al$.
For  $\I \in \Fin$
  \begin{eqnarray}
\label{eq:EO}
 \AlIe := \Ale \cap \AlI,\quad
 \AlIo := \Alo\cap \AlI.
 \end{eqnarray}
For  $A \in \Al(\AlI)$ we have the 
 even-odd decomposition:  
\begin{eqnarray}
\label{eq:EOdec}
  A=A_{+} +A_{-},\ \ 
A_{+}:= \frac{1}{2}  \bigl( A + \Theta(A)   \bigr)
\in \Ale (\AlIe),\quad 
A_{-}:=\frac{1}{2}  \bigl(A- \Theta(A)   \bigr)\in \Alo (\AlIo). 
\end{eqnarray}
 If  a pair of subsets 
$\I$ and $\J$ of $\Fin$
 are disjoint,  then 
 the following 
 graded commutation relations
 hold:
\begin{eqnarray}
\label{eq:graded}
[\AlIe,\; \AlJe]=0,\ \ [\AlIe,\; \AlJo]=0,\ \ [\AlIo,\; \AlJe]=0,
\ \ \{\AlIo,\; \AlJo\}=0,
\end{eqnarray} 
 where $[A,\ B]=AB-BA$ denotes  the commutator and 
$\{A,\ B\}=AB+BA$ the anticommutator.

We assume  that
 $\AlI$ is isomorphic to a finite-dimensional  factor
(a full matrix algebra)  
for every  $\I\in \Fin$.
Under this assumption, there is a
unitary  $\vI$ in $\AlI$
 that implements $\Theta$ on $\AlI$ as 
\begin{eqnarray}
\label{eq:vI}
   \vI^{\ast}(A)\vI =\Theta(A),\quad
 A\in \AlI.
\end{eqnarray}
This  $\vI$ is even, since 
$\Theta(\vI)=\vI^{\ast}(\vI)\vI =(\vI^{\ast}\vI)\vI =\vI$.
For disjoint $\I$ and $\J$,
the unitary $\vIJ$
of (\ref{eq:vI}) for $\AlIuJ$ 
is given by $\vI\vJ$.

The  lattice fermion system  is a typical 
example of the graded quantum systems. 
Let $\aicr$ and $\ai$  be    creation and  annihilation
 operators on the specified site $i$ in a lattice.
For each  finite subset  $\I$, 
the subsystem  $\AlI$ are 
 generated by $\aicr$ and $\ai$ in   $\I$.
The  even-odd grading transformation is  
given  by 
\begin{eqnarray*}
\label{eq:THETA}
\Theta(\aicr)=-\aicr, \quad 
\Theta(\ai)=-\ai.  
\end{eqnarray*}
The unitary  $\vI$ is given by
 $\vI := \prod_{i \in \I}\vi$, $\vi := \aicr\ai-\ai\aicr$.
We also  introduce U(1) gauge 
 transformation:
\begin{eqnarray}
\label{eq:GAUGE}
\gammat(\aicr)=e^{i \theta}\aicr, \quad 
\gammat(\ai)=e^{-i \theta}\ai\  
\end{eqnarray}
for $\theta\in \Com^{1}$.
A  state 
 invariant under $\Theta$
 is  called   even,   and that  
  invariant under $\gammat$ for 
 any 
$\theta\in \Com^{1}$
is  called   U(1)-gauge invariant.

We  provide   the plan  of this paper. 
In $\S$ \ref{sec:SSA}
we  recall  
the  strong subadditivity of entropy (SSA) 
 \cite{73lett} 
in terms of the densities  with respect to the 
 tracial state 
for general composite systems 
 made of  finite-dimensional factors.

In $\S$ \ref{sec:MARSSA},
the equivalence of 
the Markov property and the strong additivity  
 of entropy for even states of the  graded   systems is shown.
For noneven states, we have a weak result.

In $\S$ \ref{sec:MARGINAL},
we consider  restrictions of  Markov states
 onto  the marginal subsystems  that are separated from each other.
 It was shown in   \cite{HAYDEN}  \cite{MOSO}
 that  a  Markov state of 
 a three-composed tensor-product   system  
is   separable  (classically  correlated)  
 with respect to  the marginal pair of subsystems.
 We show that this   statement 
is invalid for the graded systems; 
there are U(1)-gauge invariant (hence obviously
 even) Markov states  that are
  nonseparable for  the marginal pair.
 
In $\S$ \ref{sec:IND}, we show that  a state 
 of a  graded bipartite system satisfies  
the additivity of von Neumann entropy 
 if and only if it is a product state.
This 
is  almost obvious if  the state under consideration 
is assumed to be  even.
The  point is that the evenness 
(at least on  one of the  pair of subsystems) 
follows from the assumption. 

\section{Strong subadditivity of entropy for type-I systems}
\label{sec:SSA}
We  recall  the  strong
 subadditivity  of  entropy
 for  a  general  setting that encompasses  nonindependent 
 systems.
Let $\Al$ be a 
 finite-dimensional  factor.
 Let  
 $\tau$ denote the   tracial state on $\Al$.
If an element $d\in \Al$
 is positive and  normalized as $\tau(d)=1$,
 then it is called a density.
For any state  $\vp$ of  $\Al$,
 there exists a unique  
density  $\rhovp \in \Al$ called
 the density   of  $\vp$
 satisfying that
\begin{eqnarray*}
\vp(a)=\tau(\rhovp a),\quad a\in \Al.
\end{eqnarray*}
For the tracial state $\tau$, its density is 
 obviously $\id$, the unity  of $\Al$.

Let  $\rhoone$ and $\rhotwo$ be  a pair of  densities
 of $\Al$.  
The relative entropy  for them  
 is defined  by
 \begin{eqnarray}
\label{eq:REL}
 H(\rhoone,\ \rhotwo):= 
 \tau \bigl(\rhoone (\log \rhoone  - \log \rhotwo)
\bigr) 
 \end{eqnarray}
if the support of $\rhoone$ is contained in $\rhotwo$.
Otherwise, we set it $+ \infty$.
For a pair of two 
states   $\vp$  and $\psi$ 
on $\Al$,
their  relative entropy  is   
 \begin{eqnarray}
\label{eq:RELstate}
 H(\vp,\ \psi):= H(\rho_{\vp}, \ \rho_{\psi}). 
 \end{eqnarray}

We define   the  entropy
 for  a given state $\vp$ as
\begin{eqnarray}
\label{eq:sh}
\sh(\vp) :=
 -\vp(\log \rhovp).
\end{eqnarray}
We see
\begin{eqnarray*}
\sh(\vp)=-H(\vp, \ \tau).
\end{eqnarray*} 
The following may be   
more familiar one:
\begin{eqnarray*}
S(\vp) :=
-\Tr(D_\vp\log D_\vp)= -\vp(\log D_\vp),
\end{eqnarray*}
 where $\Tr$ is 
 the matrix trace 
 that takes  $1$ for each one-dimensional projection,
and   $D_\vp$ 
denotes  the density  matrix
of $\vp$ with respect to $\Tr$.
We  see
\begin{eqnarray}
\label{eq:constant}
\sh(\vp)=S(\vp)-S(\tau)=S(\vp)-\log \Tr(\id)
\end{eqnarray} 
for any state $\vp$.
Hence if $\Al$ is  a
   $n$ by $n$  full matrix algebra, 
 $n\in\NN$, then 
 the difference $S(\vp)-\sh(\vp)$ is 
 constantly  $\log n$.

Let $\Bl$ be a subalgebra of  $\Al$.
We denote 
 the  (uniquely determined) 
conditional expectation 
 from $\Al$ onto $\Bl$ with respect to the tracial state 
 by $\EBl$.
Here,   the upper-right subscript of $E$    indicates 
 the domain and the  lower-right    
the range. 
Let $\vpBl$ denote  the restriction of $\vp$ to $\Bl$.
Then the density of ${\vpBl}$ is given by that of $\vp$ as
\begin{eqnarray}
\label{eq:densityotoshi}
\rho_{{\vpBl}}=\EBl(\rho_{\vp}).
\end{eqnarray}


%
We  have
\begin{eqnarray}
\label{eq:subtract}                    
\sh(\vpBl)-\sh(\vp)
=H(\rhovp,\  \rho_{\vpBl})
=H(\vp,\  \vpBl \otimes \tr|_{\Bex})
=H(\vp,\  \vp \circ \EBl). 
\end{eqnarray}
As a special case of (\ref{eq:subtract}),
\begin{eqnarray}
\label{eq:oritekuru}                    
\sh(\vpBl)=\sh(\vp\circ \EBl).
\end{eqnarray}

In fact we have so far assumed that 
$\vp$ is a faithful state, 
 but (\ref{eq:subtract}) is valid 
 when $\vp$ is nonfaithful.
 To see this, we  take  
$\varepsilon \cdot \tau+ (1-\varepsilon)\vp$ 
where $\varepsilon$ is a positive small number and  
then take  the limit $\varepsilon \to 0$.


Let us take three  
disjoint  subsets $A$, $B$, and $C$.
Let $\Alz, \Alab, \Albc$ and  
$\Alb$  denote finite-dimensional   quantum systems 
corresponding to the  indexes.
Let  $\Ezab$ and $\Eaba$ 
 denote the trace preserving conditional
 expectation from $\Alz$ onto $\Alab$  and that 
from $\Alab$ onto $\Ala$, respectively.
 We use similar notations    for  other indexes.
If  the domain is the total system $\Alz$,
then  we   simply  write e.g.
$\Eab$ instead of $\Ezab$ when there is no fear of confusion.

The following five conditions,
 called   the commuting square condition,
are all equivalent to each other.\\
$(${\it{1}}$)$  $\Eab  \vert_{{\Albc}}=\Ebcb$.\\
$(${\it{2}}$)$   $\Ebc  \vert_{{\Alab}}=\Eabb$.\\
$(${\it{3}}$)$ $\Alb=\Alab \cap \Albc$ and 
$\Eab \Ebc=\Ebc \Eab$.\\
$(${\it{4}}$)$ $\Eab \Ebc=\Eb$.\\
 $(${\it{5}}$)$    $\Ebc \Eab=\Eb$.\\

If our  three-composed system 
$\Alz$  satisfies the  commuting square 
 condition, then the  
  strong subadditivity of entropy $\sh(\psi)$ for any state 
$\psi$ follows.
The     proof is standard, but  
 we  recapture  it  for  completeness.
\begin{pro}
\label{pro:SSAgen}
Let $\Alz, \Alab, \Albc$ and 
 $\Alb$ be finite-dimensional  factors satisfying
 the commuting square condition, and
let $\psiz$ be an arbitrary  state on $\Alz$.
 Then
\begin{eqnarray}
\label{eq:SSAgen}
\sh(\psiz)-\sh(\psiab)-\sh(\psibc)+\sh(\psib) \leq 0.
\end{eqnarray}  
\end{pro}
Furthermore, if the system satisfies  the 
 graded commutation relations (\ref{eq:graded}),
then
\begin{eqnarray}
\label{eq:SSA}
S(\psiz)-S(\psiab)-S(\psibc)+S(\psib) \leq 0.
\end{eqnarray}  
\proof
By (\ref{eq:subtract}), (\ref{eq:oritekuru}),
and  the relation  $\Ebc \Eab  = \Eab  \Ebc  =\Eb$,
we obtain
\begin{eqnarray}
\label{eq:INE}
\sh(\psibc)-\sh(\psiz)
&=&H(\psiz,\ \psiz \circ \Ebc)\nonum\\
 &\geq& H(\psiz \circ \Eab,\ \psiz \circ \Ebc \circ \Eab)
\nonum\\
 &=& H(\psiz \circ \Eab,\ \psiz \circ \Eab \circ \Ebc) \nonum\\
 &=& H(\psiz \circ \Eab,\ \psiz \circ \Eb) \nonum\\
&=&\sh(\psiz\circ \Eb)-\sh(\psiz\circ \Eab) \nonum\\
&=&\sh(\psib)-\sh(\psiab),
 \end{eqnarray}
where the  inequality is due to the monotonicity 
 of relative entropy under the action of 
completely positive maps.

Let us turn to   the  graded systems of
 finite-dimensional  factors, which  satisfy 
the commuting square condition.
Suppose that   $\I$ and $\J$ are  disjoint subsets.
 Then the matrix trace 
on $\AlIuJ$ denoted $\TrIuJ$ is given by 
 the product extension of those in $\AlI$ and in $\AlJ$
 denoted $\TrI$ and $\TrJ$, respectively. 
Thus we have  $\TrIuJ(\id)=\TrI(\id)\times \Tr(\id)$.
 Now 
(\ref{eq:SSAgen}) and 
  (\ref{eq:constant}) imply
(\ref{eq:SSA})
\proofend\\

As this proposition indicates, the strong additivity of entropy 
 is  satisfied  irrespective of whether  states are even or not.
In \cite{CAR}
 \cite{VAL}   we have shown that noneven states may 
 have   pathological state correlations.
\section{Markov property and Strong additivity}
\label{sec:MARSSA}
It is obvious that  
the  equality of 
(\ref{eq:SSAgen}) and of (\ref{eq:SSA})
 is equivalent to that  of (\ref{eq:INE}), i.e.
\begin{eqnarray}
\label{eq:EQ}
H( \psiz,\ \psiz \circ \Ebc)
= H( \psiz \circ \Eab,\ \psiz \circ \Eb),
 \end{eqnarray}
 equivalently,
\begin{eqnarray}
\label{eq:EQsimple}
H( \rho_{\psiz},\ \rho_{\psibc})
= H( \rho_{\psiab},\ \rho_{\psib}).
 \end{eqnarray}

By  a general result of the sufficiency  given 
in \cite{PETZQUART},
  (\ref{eq:EQ}) implies  that 
the conditional expectation $\Eab$
 is sufficient for 
  $\psiz \circ \Ebc$ and $\psiz$, that is,
there exists a  completely positive map   that recovers 
 $\psiz \circ \Ebc$ and  $\psiz$ 
from  $\psiz \circ \Ebc \circ \Eab$ and 
$\psiz \circ \Eab$, respectively.
  The  canonical form of such  maps is given 
 as  follows \cite{MOSO}.

Let $\alpha$  denote the   completely positive map
 on $\Al$ defined by
\begin{eqnarray}
\label{eq:alpha}
\alpha(X) :=
{\rhopsib}^{-1/2}
\Eab \bigl( {\rhopsibc}^{1/2} X {\rhopsibc}^{1/2} \bigr)
 {\rhopsib}^{-1/2}, \quad X \in \Alz.
 \end{eqnarray}
Let $\Tsharp$ denote the dual of $\alpha$
 with respect to the tracial stare, which    
is written as 
\begin{eqnarray}
\label{eq:Tsharp}
\Tsharp(X):=
{\rhopsibc}^{1/2}
{\rhopsib}^{-1/2} 
X {\rhopsib}^{-1/2}
 {\rhopsibc}^{1/2},\quad X \in \alpha(\Alz).
 \end{eqnarray}

It is easy to see    
 $\Tsharp(\rhopsib)=\rhopsibc$. 
Also 
 $\Tsharp(\rhopsiab)=\rhopsiz$
 is satisfied if and only if 
  $\Eab$
 is sufficient for the given pair of states 
 $\psiz$ and $\psiz \circ \Ebc$, equivalently,
  (\ref{eq:EQ})  holds.

The following is a more or less   summary 
 of the  contents stated above, which 
  corresponds to Theorem 5.2 of 
 \cite{PETZMONO} and also section 5 of \cite{MOSO}
  for the tensor-product case.

\begin{pro}
\label{pro:MARSSA}
Let $\Alz, \Alab, \Albc$ and 
 $\Alb$ be finite-dimensional  factors satisfying
 the commuting square condition.  
Let $\psiz$ be an arbitrary  faithful state on $\Alz$.
The strong additivity of von Neumann 
 entropy, i.e., 
\begin{eqnarray}
\label{eq:SSAEQ}
S(\psiz)-S(\psiab)-S(\psibc)+S(\psib)=0
\end{eqnarray}  
 is satisfied  if and only if 
$\Eab$
 is sufficient for the  pair of states 
 $\psiz$ and $\psiz \circ \Ebc$. 
Let $\alpha$ denote  the  $\psiz$-preserving
(and  $\psiz \circ \Ebc$-preserving) 
 conditional expectation from $\Alz$
 to $\Alab$  given as  (\ref{eq:alpha}).
Let   $\Tsharp$ denote the dual of this $\alpha$
 with respect to the tracial state
 whose concrete formula  is given as   (\ref{eq:Tsharp}).
This $\Tsharp$ gives  the canonical left inverse
of $\Eab$  for the densities of   
$\psiz$ and $\psiz \circ \Ebc$, that is,
\begin{eqnarray}
\label{eq:rec1}
\Tsharp(\rhopsib)=\rhopsibc
\end{eqnarray}  
and 
\begin{eqnarray}
\label{eq:rec2}
\Tsharp(\rhopsiab)=\rhopsiz.
\end{eqnarray}  

The set of fixed points of  $\alpha$
 contains $\Alae$.
If the state $\psiz$ is even, then 
the set of fixed points of  $\alpha$
 contains 
$\Ala$ and  accordingly  
the    Markov property of $\psiz$ with respect to a triplet 
$(\Ala,\;\Alb,\:\Alc)$ is satisfied.
\end{pro}
\proof
We only  check    the statement   about the 
 fixed point elements of 
$\alpha$.
Take  $X\in \Alae$, which   is in the commutant 
 of $\Albc$. 
 We have 
\begin{eqnarray}
\label{eq:alphagutai}
\alpha(X) &=&
\rhopsib^{-1/2}
\Eab(\rhopsibc^{1/2} X \rhopsibc^{1/2})
 \rhopsib^{-1/2} \nonum \\
&=&\rhopsib^{-1/2}
\Eab(X \rhopsibc )
 \rhopsib^{-1/2} \nonum\\
&=&\rhopsib^{-1/2}
X \Eab(\rhopsibc) 
 \rhopsib^{-1/2}\nonum\\
&=&
\rhopsib^{-1/2}
X \Eab(\Ebc(\rhopsibc)) 
 \rhopsib^{-1/2} \nonum\\
&=&
\rhopsib^{-1/2}
X \Eb (\rhopsibc) 
 \rhopsib^{-1/2} 
\nonum\\
&=&
\rhopsib^{-1/2}
X \rhopsib 
 \rhopsib^{-1/2}
\nonum\\
&=&
X \rhopsib^{-1/2}
 \rhopsib 
 \rhopsib^{-1/2}
=X
\end{eqnarray}

Suppose now that $\psiz$ is even.
 Then $\rhopsibc \in \Albce$ 
and also $\rhopsib \in \Albe$
 commute  with 
any   $X\in \Ala$.
Hence we  see that the above set of equalities 
(\ref{eq:alphagutai})
holds for this case.
\proofend
\ \\

From this result, 
if an even state satisfies   the strong additivity of entropy, then 
the    Markov property  with respect to a triplet 
$(\Ala,\;\Alb,\:\Alc)$
 in the sense of \cite{ORI}
 (cf. Lemma 11.3 of \cite{OP}) is satisfied.

\section{Markov states on the  marginal subsystems}
\label{sec:MARGINAL}
The  definition   of separable
 states  (i.e. classically correlated  states)
for  nonindependent systems 
is  much the same as that for the tensor-product systems \cite{SEP}.
That is,
if a state is  written
 as a convex sum of some product states, 
then it is called  a separable   state.    
Let $A$ and $C$ be a pair of  disjoint subsets, 
and  $\ome$ be  a state on  $\Alac$.
If 
\begin{eqnarray}
\label{eq:EXTpair}
\ome(X Y)=\ome(X) \ome(Y)
\end{eqnarray}
for all $X  \in \Ala$ and 
 $Y  \in \Alc$, then 
 $\ome$ 
 is  called  a product state with respect to the 
 pair $(\Ala,\; \Alc)$.
It is easy to see  that the product property in  the converse order, 
\begin{eqnarray}
\label{eq:converse}
\ome(Y X)= \ome(Y)\ome(X)=\ome(X Y)
\end{eqnarray}
follows from   (\ref{eq:EXTpair}) 
and the graded commutation relations.

We discuss the property of  Markov states with respect to 
$(\Ala,\;\Alb,\;\Alc)$
for  the marginal subsystem $\Alac$. 
As we announced in the introduction, 
 Corollary 7
 of \cite{HAYDEN} is invalid for the graded systems.
\begin{pro}
\label{pro:MARGINAL}
For a three-composed graded system 
$(\Ala,\;\Alb,\;\Alc)$,
there exist  U(1)-gauge invariant  states 
 that satisfy 
the Markov property 
for 
$(\Ala,\;\Alb,\;\Alc)$
but are  nonseparable for
 $(\Ala,\;\Alc)$. 
\end{pro}
We shall  construct such Markov states.  
Using  the Jordan-Wigner transformation,  
 we set     a three-composed tensor-product 
system  in the following  way.
Let $\va$, $\vb$, and $\vab$
 denote the unitaries  implementing $\Theta$
 on the   specified subsystems. 
Let $\Alsa:=\Ala$, 
$\Alsab:=\Alab$, $\Alsz:=\Alz$,
$\Alsb:=\{\Albe, \; \va\Albo   \}$, 
 $\Alsc:=\{\Alce, \; \vab \Alco   \}$,
 and  $\Alsbc:=\{\Albce, \; \va \Albco   \}$
 where the notation `$\{\ ,\ \}$' 
denotes the algebra generated by the arguments.
They induce  a tensor-product system 
$\Alsz=\Alsa\otimes \Alsb\otimes \Alsc$.
We assign finite-dimensional Hilbert spaces 
 $\Hilspa$, $\Hilspb$  and $\Hilspc$
 to $\Alsa$, $\Alsb$ and $\Alsc$, respectively.
We will use the  following  lemma   
 later. Its  proof  is   obvious.
\begin{lem}
\label{lem:uturu}
Let $\psiz$ be an arbitrary even  state on $\Alz$.
It satisfies 
\begin{eqnarray}
\label{eq:uturucar}
S(\psiz)-S(\psiab)-S(\psibc)+S(\psib) = 0,
\end{eqnarray}  
if and only if
\begin{eqnarray}
\label{eq:uturutensor}
S(\psiz)-S(\psi|_{\Alsab})-S(\psi|_{\Alsbc})+S(\psi|_{\Alsb}) 
= 0.
\end{eqnarray}  
\end{lem}
\ \\

For a while we will focus on
 the two composed system, $\Alac$.   
In \cite{SEP}
we have discussed  
 how  the  state correlation (separability,  
 nonseparability) will remain or change
 under the Jordan-Wigner transformation 
 which maps   the CAR pair $(\Ala, \;\Alc)$  
 to $(\Ala, \;\Alscac)$,  
 where   $\Alscac$ denotes  the commutant of $\Ala$
 in $\Alac$ 
and is explicitly given as   
$\{\Alce, \; \va \Alco   \}$. 
(Note that $\Alscac$ is different from previously 
introduced $\Alsc$.)
It has  been  shown  that 
 the set of all separable states for the CAR
 pair  is strictly smaller than
 that for the tensor-product pair. That is,  
if $\omeac$ is a separable state  for the pair 
 $(\Ala, \Alc)$, then so it is for 
 $(\Ala, \Alscac)$. However,
 there exist U(1)-invariant  states that 
 are separable for the latter  but 
 nonseparable for the former.
We introduce    an  
 example of such states from  \cite{SEP}.

\newcommand{\Ka}{k_{A}}%
\newcommand{\Kc}{k_{C}}%
\newcommand{\Kad}{k^{\ast}_A}%
\newcommand{\Kcd}{k^{\ast}_C}%
\newcommand{\hopterm}{1/2(\Kad \Kc-\Ka \Kcd)}

Let $\Ka$ and  $\Kc$ be some nonzero odd elements 
in $\Ala$ and in $\Alc$, e.g. field operators on 
specified regions.  
Let  $K:=\hopterm$ which is self-adjoint and denotes 
fermion-hopping interaction between $\Ala$ and $\Alc$.
Suppose that $\Vert \Ka \Vert\le 1$
and   $\Vert \Kc  \Vert\le 1$, 
 then   $\Vert K  \Vert\le 1$.
For $\lam\in \R$, $|\lam|\le 1$,  
$\rhoaclam:= \id+ \lam K$ gives a 
density  operator. 
 For $0<|\lam|\le 1$,
 the state on $\Alac$ with its  density 
$\rhoaclam$ gives a  state satisfying  
all the  desired conditions.

Now take   such a U(1)-gauge invariant 
state $\omeac$  on $\Alac$.
 It has  a   state-decomposition
$\omeac =\sum_{i=1}^{n} \lami  \omeaci$, 
$0<\lami< 1$, $\sum \lami=1$,
such that 
each $\omeaci$ is a  product state  
for  $(\Ala, \; \Alscac)$, but   has  no product-state 
decomposition for $(\Ala, \Alc)$.
 From this,
we are going to construct   a state on $\Alz$ that proves Proposition 
\ref{pro:MARGINAL}.

Let us assume that the dimension of 
$\Hilspb$ is equal or more than $n$.
 Then we have a set of $n$ nonzero 
even orthogonal projections 
$p_i\in \Albe$,  $1\le i\le n$. 
Let $\omebi(X):=\tau(p_i X)/\tau(p_i)$,
 for $X\in \Alb$. Those are all   even states of $\Alb$.
Let $\omeabc:= \sum_{i=1}^{n} \lami  \omeaci \circ \omebi$,
 where $\omeaci \circ \omebi$ denotes the 
(uniquely determined) product state extension of 
$\omeaci$ on $\Alac$ and $ \omebi$ on $\Alb$, see \cite{AM}.

We will see that $\omeaci \circ \omebi$
 gives a  product state 
for $(\Ala,\;\Alsc)$ when restricted to $\Ala\otimes \Alsc$.
We must   check this 
 for the product element 
$a\cev$ such that 
 $a\in \Ala$
 and $\cev\in \Alce$, 
and for $a(\va\vb\cod)$ such that 
  $a\in \Ala$
 and $\cod\in \Alco$. 
We have 
\begin{eqnarray}
\label{eq:ACspinproduct1}
 \omeaci \circ \omebi(a\cev)=\omeaci(a\cev)=
\omeaci(a)\omeaci(\cev)=
\omeaci \circ \omebi(a)
\omeaci \circ \omebi
 (\cev),
\end{eqnarray}  
and using the  product property of $\omeaci$  
for  $(\Ala, \; \Alscac)$ 
\begin{eqnarray}
\label{eq:ACspinproduct2}
 \omeaci \circ \omebi(a \va\vb\cod)&=&
 \omeaci \circ \omebi(a \va\cod\vb)\nonum\\
&=& \omeaci(a \va\cod) \omebi(\vb)\nonum\\
&=& \omeaci(a) \omeaci(\va\cod) \omebi(\vb)\nonum\\
&=& \omeaci (a)
\omeaci \circ \omebi
 (\va\cod \vb)\nonum \\
&=&\omeaci \circ \omebi(a)
\omeaci \circ \omebi
 (\va\vb\cod).
\end{eqnarray}  
Hence  
 $\omeaci \circ \omebi$
 has  a  product state restriction, 
 and accordingly $\omeabc$ has  a separable 
 state restriction  for   $(\Ala,\;\Alsc)$.
We conclude that our  $\omeabc$
  has the structure 
as  in   Theorem 6 of \cite{HAYDEN}
 or as the formula (14)  of \cite{MOSO}
with respect to   ($\Hilspa$, $\Hilspb$, $\Hilspc$).
Hence
it
satisfies the  Markov property  with respect to  
$(\Alsa,\;\Alsb,\;\Alsc)$.
 
From the equivalence of the Markov property
 and the strong additivity of entropy for  three-composed 
tensor-product  systems, which has been  shown 
 in the above  references,
(\ref{eq:uturutensor}) is satisfied  for  $\omeabc$.
Since it is even,   it satisfies 
 (\ref{eq:uturucar})  as well and hence 
is Markovian   
with respect to  
$(\Alsa,\;\Alsb,\;\Alsc)$ by 
Proposition \ref{pro:MARSSA}.
As  $\omeac|_{\Alac}=\omeac$    is obviously nonseparable
 for $(\Ala,\;\Alc)$  by  definition, 
$\omeabc$  gives a state showing  
 Proposition \ref{pro:MARGINAL}.


\section{Additivity of von Neumann entropy and the product 
 property}
\label{sec:IND}
We consider 
 a two-composed graded system $\Alac$ generated 
 by $\Ala$ and $\Alc$.
 Namely, we treat   the case where the intersection 
 region  $B$ is empty.
Then the strong subadditivity 
of entropy (\ref{eq:SSA})
becomes
\begin{eqnarray}
\label{eq:SA}
S(\psiac)-S(\psia)-S(\psic) \le  0,
\end{eqnarray}  
 which is called   the subadditivity of entropy.
We discuss  
 characterization of additivity of entropy, i.e. the 
 condition of equality of this inequality.
 
The answer is   very simple  for 
 tensor-product systems: 
 a  state satisfies the additivity of entropy 
 if and only if it is a product state.
For the graded system,
 we can  show  a similar result easily
under the assumption that  
the marginal states 
$\psia$ and $\psic$
are not both noneven. 
Let $\psia\circ\psic$ denote the product state of 
 $\Alac$ whose restrictions  to $\Ala$ and $\Alc$ are  
 $\psia$  and $\psic$.  Its existence 
 is guaranteed if $\psia$ or-and $\psic$ is even. 
Then we have 
\begin{eqnarray}
\label{eq:SAeq}
S(\psiac)-S(\psia)-S(\psic)= -H(\psiac, \psia\circ\psic)\le 0.
\end{eqnarray}  
By the strict  positivity  of relative entropy, it is $0$
 if and only if $\psiac=\psia\circ\psic$.

Now we drop  the  evenness  assumption  
 on  the  states.
If $\psia$  and $\psic$ are both  noneven, then 
there is  no  product state extension for them \cite{AM}.
 Hence the above argument  using the strict positivity
 of relative entropy does not work for the general   case. 

Using     
 \cite{MOSO} we derive    
 the  following.
\begin{pro}
\label{pro:IND}
Let $\psiac$ be a state of the two-composed graded system 
$\Alac$.
 It satisfies the additivity of entropy
\begin{eqnarray}
\label{eq:ADD}
S(\psiac)-S(\psia)-S(\psic)= 0,
\end{eqnarray}  
if and only if 
it is a product state for
 $(\Ala, \;\Alc)$. 
If it is the  case, at least one of $\psia$
and $\psic$ is even.
\end{pro}
\proof
The equivalence of  (\ref{eq:ADD})
and  (\ref{eq:EQsimple}) when
 the middle part $B$ is empty implies that  (\ref{eq:ADD})
 is equivalent  to  
\begin{eqnarray}
\label{eq:EQsimpleII}
H( \rho_{\psiac},\ \rho_{\psic})
= H( \rho_{\psia},\ \id).
 \end{eqnarray}
This is equivalent to say that 
 $\Ea$
 is sufficient for $\rho_{\psiac}$ and $\rho_{\psic}$.
\newcommand{\psicsq}{\rho^{\ 1/2}_{\psic}}
\newcommand{\psiasq}{\rho^{\ 1/2}_{\psia}}
\newcommand{\psiacsq}{\rho^{\ 1/2}_{\psiac}}
Now from (\ref{eq:Tsharp})
the canonical left inverse 
 of  $\Ea$ for those densities is given by
 \begin{eqnarray}
\label{eq:Tsharpbnashi}
\Tsharp(X):=
{\psicsq} X {\psicsq},\quad X \in  \Ala.
 \end{eqnarray}
Hence  we have 
\begin{eqnarray}
\label{eq:recac}
\rhopsiac=  \Tsharp(\rhopsia)=\psicsq\rhopsia\psicsq.
\end{eqnarray}  
Exchanging  $A$ and $C$ and repeating  the same argument 
as above,
 we  have also   
\begin{eqnarray}
\label{eq:recca}
\rhopsiac= \psiasq
\rhopsic
\psiasq.
\end{eqnarray}  

Let us take  the  
decomposition of $\rhopsia$ 
 into its even-odd parts and   that of $\rhopsic$ as  
 in (\ref{eq:EOdec}):
\begin{eqnarray}
\label{eq:decrhopsia}
  \rhopsia&=&\rhopsiae+\rhopsiao, \quad \rhopsiae\in \Alae,
\  \rhopsiao \in \Alao,\nonum \\
  \rhopsic&=&\rhopsice+\rhopsico, \quad \rhopsice\in \Alce,
\  \rhopsico \in \Alco.
 \end{eqnarray}
Similarly  take  the
 even-odd   
decomposition of $\psiasq$  and   that of $\psicsq$ 
 in  the following:
\begin{eqnarray}
\label{eq:decpsiasq}
  \psiasq&=&\aev+\aod, \quad \aev\in \Alae,\  \aod \in \Alao,\nonum \\
  \psicsq&=&\cev+\cod, \quad \cev\in \Alce,\  \cod \in \Alco. 
 \end{eqnarray}
Since the densities are positive hence  self-adjoint,
 each of  $\aev$, $\aod$,  $\cev$, and $\cod$
 is self-adjoint.
We have 
\begin{eqnarray}
\label{eq:termsI}
  \rhopsia&=& (\psiasq)^{2}
=\aev^{2}+\aod^{2} +\aev\aod+\aod \aev,\nonum\\
  \rhopsiae&=& \aev^{2}+\aod^{2}, \nonum\\
  \rhopsiao&=& \aev\aod+\aod \aev,
\end{eqnarray}
 and 
\begin{eqnarray}
\label{eq:termsI}
  \rhopsic&=&\cev^{2}+\cod^{2} +\cev\cod+\cod \cev,\nonum\\
  \rhopsiae&=& \cev^{2}+\cod^{2}, \nonum\\
  \rhopsiao&=& \cev\cod+\cod\cev.
\end{eqnarray}
Now we shall express
 the equality
$\psicsq\rhopsia\psicsq
=\psiasq\rhopsic\psiasq =
\rhopsiac$ 
in terms of  $\aev$, $\aod$,  $\cev$ and $\cod$.
We compute
\begin{eqnarray}
\label{eq:cac}
&&\psicsq\rhopsia\psicsq \nonum\\
&=&
\psicsq(\rhopsiae+\rhopsiao )\psicsq\nonum\\
&=& \left(\rhopsiae \psicsq   +\rhopsiao \Theta\left(\psicsq\right) 
\right)\psicsq =
\rhopsiae \rhopsic   
+\rhopsiao \Theta\left(\psicsq \right)\psicsq\nonum\\
&=& 
(\aev^{2}+\aod^{2})(\cev^{2}+\cod^{2} +\cev\cod+\cod\cev)
  +  (\aev\aod+\aod\aev)  (\cev-\cod)(\cev+\cod) \nonum\\ 
&=& 
\aev^{2}(\cev^{2}+\cod^{2} +\cev\cod+\cod\cev)
+\aod^{2}(\cev^{2}+\cod^{2} +\cev\cod+\cod\cev)\nonum\\
&&+ \aev\aod (\cev^{2}-\cod^{2}-\cod\cev+\cev\cod)+
\aod\aev  (\cev^{2}-\cod^{2}-\cod\cev +\cev\cod). 
\end{eqnarray}
Also,
\begin{eqnarray}
\label{eq:aca}
&&\psiasq\rhopsic\psiasq \nonum\\
&=&
\psiasq(\rhopsice+\rhopsico )\psiasq\nonum\\
&=& 
\rhopsia \rhopsice+ \psiasq \Theta\left( \psiasq  \right)
\rhopsico
\nonum\\
&=&
(\aev^{2}+\aod^{2} +\aev\aod+\aod\aev)
(\cev^{2}+\cod^{2})+
(\aev^{2}-\aod^{2} -\aev\aod+\aod \aev)
(\cev\cod+\cod\cev),\nonum\\
 &=&
\aev^{2}(\cev^{2}+\cod^{2} +\cev\cod+\cod\cev)+
\aod^{2}(\cev^{2}+\cod^{2} -\cev\cod-\cod\cev)\nonum\\
&&\quad +\aev\aod(\cev^{2}+\cod^{2} -\cev\cod-\cod\cev)
+ \aod \aev (\cev^{2}+\cod^{2} +\cev\cod+\cod\cev).
 \end{eqnarray}
Equating (\ref{eq:cac})
and (\ref{eq:aca}), we have
\begin{eqnarray}
\label{eq:equating}
\aod^{2}(\cev\cod+\cod\cev)+\aev\aod(-\cod^{2}+ \cev\cod )+
\aod\aev(-\cod^{2}-\cod\cev )=0.
 \end{eqnarray}
Taking  the  even and odd parts of this, we have
\begin{eqnarray}
\label{eq:equatinge}
\aev\aod\cev\cod -\aod\aev\cod\cev =0,\\
\label{eq:equatingo}
\aod^{2}(\cev\cod+\cod\cev) 
-(\aev\aod +\aod\aev)  \cod^{2} =0.
 \end{eqnarray}
By acting
 the unitary transformation  
Ad$(\vA)$ on the both sides of 
(\ref{eq:equatingo})
 where   $\vA$ in $\Alae$
 gives  the implementation of  $\Theta$ on $\Ala$ as 
(\ref{eq:vI}), we have
\begin{eqnarray*}
\label{eq:}
\aod^{2}(\cev\cod+\cod\cev) 
+(\aev\aod +\aod\aev)  \cod^{2} =0.
 \end{eqnarray*}
By  averaging this and (\ref{eq:equatingo}),
 we have
\begin{eqnarray}
\label{eq:equatingoI}
\aod^{2}(\cev\cod+\cod\cev)=0. 
 \end{eqnarray}
Similarly, we have 
\begin{eqnarray}
\label{eq:equatingoII}
(\aev\aod +\aod\aev)  \cod^{2} =0.
 \end{eqnarray}

We will see that 
from  (\ref{eq:equatinge}), (\ref{eq:equatingoI})
 and (\ref{eq:equatingoII}),
 our assertion, i.e. the eveness of  
 $\rhopsia$ or (and) $\rhopsic$
follows.  
For   (\ref{eq:equatingoI}) to be satisfied,
\begin{eqnarray}
\label{eq:equatingoIor}
\aod^{2}=0 \quad  {\mbox{or-and}}  \quad (\cev\cod+\cod\cev)=0, 
 \end{eqnarray}
as  $\aod^{2}\in \Alae$ and hence 
$\aod^{2}(\cev\cod+\cod\cev)=\aod^{2}\otimes(\cev\cod+\cod\cev)=0$. 
In the same way,
\begin{eqnarray}
\label{eq:equatingoIor}
\cod^{2}=0 \quad {\mbox{or-and}} \quad (\aev\aod+\aod\aev)=0. 
 \end{eqnarray}
If $\aod^{2}=0$, then $\aod=0$ 
 since $\aod$ is self-adjoint.
Therefore  $\psiasq$ is even and  so $\rhopsia$ is.
If $\cod^{2}=0$, then  $\rhopsic$ is even.
We now consider the remaining possibility, i.e. the case where  
$\aev\aod+\aod\aev=\cev\cod+\cod\cev=0$.
 This implies that $\rhopsiao=\rhopsico=0$,
 namely both of $\rhopsia$ and $\rhopsic$ are even.
In conclusion,  
 at least one of the marginal states  
$\rhopsia$ and $\rhopsic$ should be even.

Now we know that 
the product state  $\psia\circ\psic$ exists 
 and can use the argument 
 in  (\ref{eq:SAeq}) that  leads to  our  desired assertion. 
\proofend
\ \\

We shall go back to   three-composed systems 
and comment on the   condition of 
the strong additivity of entropy.
For now,  we are only   able to produce the desiarble 
 form of Markov property for even states.
We guess that the  strong additivity  
of entropy may control  {\it{nonevenness}}
 of the  states   
 as   for  the special 
case of  two-composed systems.
\ \\
\ \\
{\it{Acknowledgements.}}\ \\
The author  thanks   paticipants 
 of   the  
von Neumann conference, Budapest, 
October 2003 and   of 
26th  conference of  QP and IDA
 in  Levico (Trento), February, 2005 for useful conversation,
especially
 L. Accardi and D. Petz  for  explanation on    their works.

\end{document}